\documentclass[times]{ettauth}

\usepackage{flushend}
\usepackage{lipsum}
\usepackage{amsmath}
\usepackage{mathtools}
\usepackage{cuted}
\usepackage{cite}
\usepackage[T1]{fontenc}
\usepackage{amssymb}
\usepackage{graphicx}
\usepackage{sidecap}
\usepackage{bigints}\usepackage{moreverb}

\usepackage[colorlinks,bookmarksopen,bookmarksnumbered,citecolor=red,urlcolor=red]{hyperref}

\newcommand\BibTeX{{\rmfamily B\kern-.05em \textsc{i\kern-.025em b}\kern-.08em
T\kern-.1667em\lower.7ex\hbox{E}\kern-.125emX}}

\def\volumeyear{2017}

\begin{document}

\runningheads{S. ~Ghosh ~et. ~al.}{Outage analysis in two-way communication with RF energy harvesting ..}

\articletype{RESEARCH ARTICLE}

\title{Outage analysis in two-way communication with RF energy harvesting relay and co-channel interference}

\author{Sutanu ~Ghosh$^{\dagger}$, Tamaghna ~Acharya\corrauth$^{\dagger}$ and Santi ~P. ~Maity$^{\ddagger}$}
\address{$^{\dagger}$Department of Electronics and Telecommunication Engineering,\\
$^{\ddagger}$Department of Information Technology,\\
Indian Institute of Engineering Science and Technology,
Shibpur, Howrah-711103, India}

\corraddr{Tamaghna Acharya,
Department of Electronics and Telecommunication Engineering, 
Indian institute of engineering science and technology, Shibpur, West Bengal, India, 
Email : t\char`_acharya@telecom.iiests.ac.in
}

\begin{abstract}
The study of relays with the scope of energy-harvesting (EH) looks interesting as a means of enabling sustainable, wireless communication without the need to recharge or replace the battery driving the relays. However, reliability of such communication systems becomes an important design challenge when such relays scavenge energy from the information bearing RF signals received from the source, using the technique of simultaneous wireless information and power transfer (SWIPT). To this aim, this work studies bidirectional communication in a decode-and-forward (DF) relay assisted cooperative wireless network in presence of co-channel interference (CCI). In order to quantify the reliability of the bidirectional communication systems, a closed form expression for the outage probability of the system is derived for both power splitting (PS) and time switching (TS) mode of operation of the relay. Simulation results are used to validate the accuracy of our analytical results and illustrate the dependence of the outage probability on various system parameters, like PS factor, TS factor, and distance of the relay from both the users. Results of performance comparison between PS relaying (PSR) and TS relaying (TSR) schemes are also presented. Besides, simulation results are also used to illustrate the spectral-efficiency and the energy-efficiency of the proposed system. The results show that, both in terms of spectral-efficiency and the energy-efficiency, the two-way communication system in presence of moderate CCI power, performs better than the similar system without CCI. Additionally, it is also found that PSR is superior to TSR protocol in terms of peak energy-efficiency.
\end{abstract}

\maketitle
\footnotetext[2]
{\tiny%
\href{http://onlinelibrary.wiley.com/journal/10.1002/(ISSN)2161-3915}{\texttt{http://onlinelibrary.wiley.com/journal/10.1002/(ISSN)2161-3915}}
}

\section*{Keywords}
{Two-way communication; simultaneous wireless information and power transfer; decode-and-forward relaying.}

\section{Introduction}
Recently, relay assisted communication enjoys extensive popularity in wireless networks in view of its ability to improve network performance in terms of improving its throughput, reliability, coverage, energy-efficiency, interference mitigation etc. However, relay nodes are mostly battery driven. It may not always be feasible to drive the relays by grid power or recharge their batteries whenever necessary. This poses a serious challenge in terms of network lifetime and flexibility of deployment. Energy harvesting (EH) [1] technology is fast emerging as a promising solution to the aforementioned problem, where a low power node like a relay, equipped with necessary hardware would be capable of harvesting energy from various sources present in the surrounding environment like solar, wind, radio frequency (RF) signals [2] etc. Recently, EH from RF signal has attracted a lot the attention of research community. Success of the RF-EH technology would be a major boost to the growing applications of Internet of Things (IoT) [3], in terms of improving lifetime of the IOT devices  without the need of recharging or replacing their batteries. Recent literature report that theoretical maximum power available for RF-EH at a free space distance of 40 meters is 7 $\mu$W and 1 $\mu$W for 2.4 GHz and 900 MHz frequency, respectively [4]. Two popular techniques, found in open literature, for RF-EH in wireless networks  are : (i) wireless powered communication network (WPCN) [5] and (ii) simultaneous wireless information and power transfer (SWIPT) [6], [7]. While the former refers to the process of RF-EH from base station in downlink to power the uplink information transfer by the corresponding wireless device, the latter describes a technique for RF-EH from the RF signal originally meant for information transfer. In this paper, we focus on the latter one.

Two different receiver architectures are proposed for SWIPT; power-splitting (PS) and time-switching (TS) [6]. In PS protocol, receiver is capable to split the received power from the RF signal into two parts: one to scavenge energy and the remaining part to process information. In TS scheme, in each transmission frame, EH and information decoding (ID) are performed in two adjacent and non-overlapping time slots.

RF-EH has an important role in cooperative relay network. Based on the energy causality
constraint, the usable energy of relay cannot exceed the energy harvested by it. This imposes major
limitations on the system performance at the end-to-end link. So, the efficient usage of RF-EH relay in various forms of cooperative wireless networks is studied in [8-16], [19]. Performances of these networks are evaluated in terms of outage probability and throughput in both one-way [8-12] and two-way [13-15] communications. In [8], amplify-and-forward (AF) relay assisted and in [9], decode-and-forward (DF) relay assisted PS relaying (PSR) and TS relaying (TSR) schemes are studied for delay-limited and delay-tolerant traffic, respectively. The authors of [8] report that TSR performs better than PSR in terms of throughput at low signal-to-noise ratio (SNR) and high transmission rate in AF system. On the other hand, the performance of PSR is better than TSR in DF system for a wide range of SNR [9]. Sanjay et al. [10] analyse an AF relay assisted model using TSR scheme to show the impact of relay placement in cognitive radio network. The work in [11] makes a comparative study between AF and DF relays, in terms of outage probability in PS protocol architecture. Furthermore, the performance of both PSR and TSR schemes in two-hop relay-assisted multi-input multi-output (MIMO) system are analyzed in nakagami fading channel [12].

To enhance the spectrum efficiency, two-way communication using relaying scheme is more useful than two parallel one-way communication. The relay placement is one of the important aspect of research in two-way communication for harvesting an useful amount of energy from both of the users [15]. Two-way communications using AF relay assisted architecture is discussed in [13], [15]. In [13], finite expressions of outage probability and ergodic capacity are presented for PSR scheme. In [14], a closed form of outage probability is established in DF relay assisted two-way communication with PSR technique. On the other hand, power-time splitting  based hybrid architecture in two-slot and three-slot relaying protocols are discussed in [15]. The work reports the system performance in terms of outage probability and throughput in delay-limited and delay-tolerant transmissions.  

Some studies on resource optimization in RF-EH enabled one-way and two-way communications are also reported in [14],[16-17]. In [16], outage minimization is done with respect to harvesting power using classical optimization technique. On the other hand, due to non-convex nature of outage minimization problem in [14], a meta heuristic approach like genetic algorithm (GA) base complex search technique is used to solve the optimization problem. Optimal policies of dynamic power splitting technique is proposed in [17], based on the available and unknown channel state information (CSI) at the transmitter. In their work, maximization of ergodic capacity of information transmission is shown for a given harvesting energy constraint. 

Interference is traditionally considered to be one of the major challenges in wireless communications due to its adversing effect on throughput. The impact of co-channel interference (CCI) in two-hop, multiple AF relay assisted system is studied in [18] to compare the optimum combining and maximal ratio combining performance. The presence of CCI in SWIPT enabled wireless communication is expected to have an intriguing impact on the system performance. On one hand, it can be considered as a useful source to scavenge additional energy through RF-EH technique, on the other hand, it acts as a deterrent to information decoding process at the receiver. This dual yet conflicting role of CCI is investigated in [19]. The authors report SWIPT enabled, single DF relay assisted one-way communications system's performance in terms of its outage probability and ergodic capacity for both PSR and TSR schemes. To the best of knowledge, the impact of CCI in two-way DF relay assisted network is not yet explored and this is the motivation of our current study.
   
In this paper, bi-directional DF relay assisted communication is considered for PSR (TSR) with frame structure consists of a three slots (four slots). The users (data and power transmission nodes) are driven by their individual unlimited energy resources, whereas due to lack of any steady energy resource, the relay node is harvesting energy only from the RF signals received from the users by means of SWIPT technique. The harvested energy is used to support data transfer between two users. Additionally, the presence of CCI is considered as a source of EH. A closed form expression of outage probability is derived for PSR and TSR schemes. Numerical results are used to compare the system performance between two relaying protocols of SWIPT. Further, their performance results are also compared with a similar network architecture without considering the presence of CCI. This would help to gain insight the relative contributions of CCI in such networks. 

A list of symbols and their definitions are shown in Table I for better readability. Remaining part of this paper is organized as follows. System model is described in Section II. Outage analysis for the two way relaying protocol is presented in Section III and the performance evaluation is reported in Section IV. Finally, the paper is concluded in Section V.

\begin{table}\caption{Symbols and definitions}
  \centering
    \begin{tabular}{l|p{50mm}}
    \hline
    Symbols & Definitions\\
    \hline
    \hline
     $ s_{k}^{PS}$ & Signal of $User_k$ in PSR\\
    \hline
     $ s_{k}^{TS}$ & Signal of $User_k$ in TSR \\  
    \hline
     $s_{j}^{PS}$ & Signal of interferer in PSR \\  
    \hline
     $s_{j}^{TS}$ & Signal of interferer in TSR \\
     \hline
     $n_{r,at_k}$ & Antenna noise\\
    \hline
     $n_{r,sc}$ & RF to baseband signal conversion noise \\
    \hline
     $P_{k}^{T}$  &  Transmission power $User_k \rightarrow$ {\em EHR}\\
    \hline
     $P_{k}^{R}$  &  Receive power at {\em EHR}\\
    \hline
     p  &  CCI power\\
    \hline
    $d_k$ &  Distance between $User_k$ and {\em EHR}\\
    \hline
    $\eta$ & Energy conversion efficiency\\
    \hline
    $\nu$ &  Path loss exponent\\
    \hline
    $h_k$  & Channel gain of link $User_k \rightarrow$ {\em EHR}\\
    \hline
    $g_k$  & Channel gain of link {\em EHR} $\rightarrow User_k$ \\
    \hline
    $\beta_c$  & Channel gain of link Interferer $\rightarrow$ {\em EHR}\\
    \hline
    $\alpha_k (k\in1,2)$  &   Power splitting factor\\
    \hline
    $\rho_k (k\in1,2)$  &  Time switching factor\\
    \hline
    $R_{k,r}^{PS}$ & Data rate $User_k \rightarrow$ {\em EHR} in PSR \\
    \hline
    $R_{k,r}^{TS}$ & Data rate $User_k \rightarrow$ {\em EHR} in TSR \\
    \hline
     $R_{rs_k}^{PS}$ & Data rate {\em EHR} $ \rightarrow  User_k$ in PSR \\
    \hline
    $R_{rs_k}^{TS}$ & Data rate {\em EHR} $ \rightarrow  User_k$ in TSR \\
    \hline
    $R_{bcrs}^{PS}$  &  Relay broadcast rate in PSR\\
    \hline
    $R_{bcrs}^{TS}$  &  Relay broadcast rate in TSR\\
    \hline
    $P_{out, U_{k}R}^{PS}$  &  Outage probability of link $User_k \rightarrow$ {\em EHR} in PSR\\
    \hline
    $P_{out, U_{k}R}^{TS}$  &  	Outage probability of link $User_k \rightarrow$ {\em EHR} in TSR\\
    \hline
    $P_{out,BC}^{PS}$  &  Outage probability due to relay broadcast in PSR\\
    \hline
    $P_{out,BC}^{TS}$  &  Outage probability due to relay broadcast in TSR\\
    \hline
    $P_{out}^{PS}$  &  Outage probability of proposed system in PSR\\
    \hline
    $P_{out}^{TS}$  &  Outage probability of proposed system in TSR\\
   \hline
   $\gamma_{h_k}^{PS}$   &  SNR of link $User_k \rightarrow$ {\em EHR} in PSR\\
   \hline
   $\gamma_{h_k}^{TS}$   &  SNR of link $User_k \rightarrow$ {\em EHR} in TSR\\
   \hline
   $I_{R_k}^{PS}$   &  Interference to noise ratio in PSR\\
   \hline
   $I_{R_k}^{TS}$   &  Interference to noise ratio in TSR\\
   \hline
         \end{tabular}
  \end{table}

\section{System Model and Protocol Description}

This section describes the system model in the framework of PSR and TSR protocols.

\subsection{System Model}

\begin{figure}
\centering
{\includegraphics[scale=0.45]{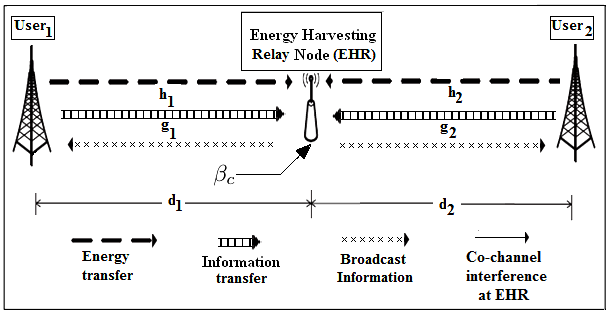}
\caption {System Model}  \label{fig:img_1}}
\end{figure}

System model considered in this work as illustrated in Figure 1, consists of half-duplex, two-way, cooperative DF relaying system, where two users - $User_1$ and $User_2$ can communicate with each other through an intermediate EH relay {\em(EHR)} node. Data transmission from each user is done in two adjacent, non-overlapping time slots. Both the users and the relay use single antenna. The users are powered by conventional power resources which are assumed to be unlimited. However, the relay is assumed to be driven solely by the harvested energy from the received RF signals of the users, following the principle of SWIPT. The links between $User_k$ and {\em EHR} is considered to be Quasi-static fading to characterize all the independent channel coefficients, which are complex circularly symmetric Gaussian random variables and remain fixed over the interval of two way transmission time \textit{T}. The channel coefficients of the links between $User_{k}$ to {\em EHR} and {\em EHR} to $User_{k}$ ($k \in 1, 2$) are represented by ${h_k}$ and ${g_k}$, respectively. Additional impact of CCI is also considered due to inherent frequency reuse for efficient utilization of the spectrum resource. The link between interferer and {\em EHR} is assumed to be  slow-fading [18], so the channel gain $\beta_{c}$ remains constant during single time frame. Further, their individual distributions are described as $h_{k} \sim {\mathcal{CN}}(0,\Omega_{h_k})$, $g_{k} \sim {\mathcal{CN}}(0,\Omega_{g_k})$  and $\beta_{c}\sim {\mathcal{CN}}(0,\Omega_{\beta_c})$ ($k \in 1, 2$), where $0$ is the mean value and $\Omega_{h_k}, \Omega_{g_k}, \Omega_{\beta_c}$ are the variances of channel coefficients $h_k, g_k, \beta_{c}$, respectively.   

  Following [18], it is assumed that the necessary channel state information (CSI) is available for the given system model in CCI environment. The signal received at {\em EHR} from $User_{k}$ can be expressed as 

\begin{equation}
 \centering
P_{k}^{R}={\scalebox{1.2} {$\frac{P_{k}^{T}|h_{k}|^2}{(d_{k})^\nu}$}}, k \in 1,2
\end{equation}
where, the symbols $P_{k}^{T}$, $\nu$ and $d_{k}$ are represented as the transmission power, path loss exponent and the distance between $User_{k}$ to {\em EHR}, respectively. 

\subsection{PS relaying scheme}

\begin{figure}
\centering
\includegraphics[scale=0.64]{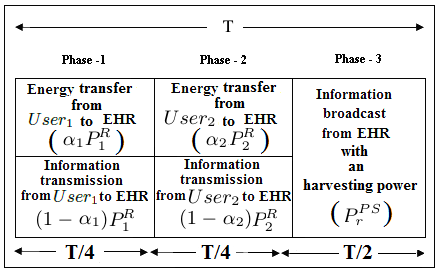}
\caption {Transmission Framework for PSR}  \label{fig:img_2}
\end{figure}

 The bi-directional communication process takes place on a frame-by-frame basis. Duration of any such frame is considered $T$, splits into three time slots : $\dfrac{T}{4}$, $\dfrac{T}{4}$ and $\dfrac{T}{2}$ as shown in Figure 2. During the first two slots, $User_1$ and $User_2$ send their individual information to {\em EHR} in their corresponding slots, where total received power from each user is split into two parts using PS protocol. The splitting ratio of the received power $ P_{r_k} $  is $ \alpha_{k} $:$ (1-\alpha_{k}) $, where 0 $<$ $ \alpha_{k} $ $<1$. First part of received power $ \alpha_{k}P_{r_k} $ is used for information decoding and the rest $ (1-\alpha_{k})P_{r,k} $ is used for EH. Finally in Slot 3, after successful information decoding from the messages of both the users, {\em EHR} re-encodes them and broadcasts signals to $ User_{1} $ and $ User_{2} $. In this stage of encoding, {\em EHR} applies the digital network coding (NC) (like, XOR operation) to mix the received information of $ User_{1} $ and $ User_{2} $. Finally, $ User_{1} $ and $ User_{2} $ can decode the desired information from the mixed information using the standard principle followed in NC-based relaying protocol [20]. 

\subsection{TS Relaying scheme}

\begin{figure}
\includegraphics[scale=0.565]{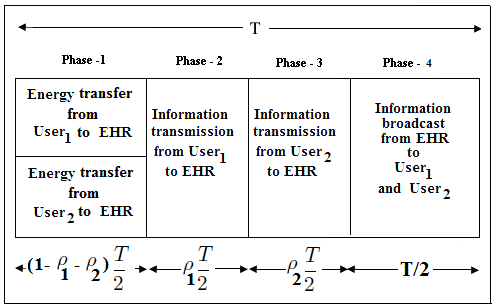}
\caption {Transmission Framework for TSR}  \label{fig:img_3}
\end{figure}

Similar to PS in the section 2.2, the bidirectional communication is held on a frame-by-frame basis. 
As shown in Figure 3, total frame duration {\em T} is split into four different slots : $\rho_1\dfrac{T}{2}$, $\rho_2 \dfrac{T}{2}$,  $(1-\rho_1-\rho_2)\dfrac{T}{2}$ and  $\dfrac{T}{2}$; where $\rho_1$, $\rho_2$ are time splitting factors and 0 $<$ $\rho_1+\rho_2$ $<$1. During the first slot of time frame $(1-\rho_1-\rho_2)\dfrac{T}{2}$, {\em EHR} harvests energy from $User_1$ and $User_2$. In the second slot $\rho_1 \dfrac{T}{2}$, $User_1$ transmits its information to {\em EHR}. Similar task is performed by $User_2$ in the third slot $\rho_2 \dfrac{T}{2}$. Similar to PS protocol, in the final slot {$\dfrac{T}{2}$}, after successful information decoding from the message of $ User_{k} $, {\em EHR} broadcasts a signal to $ User_{1} $ and $ User_{2} $. After reception, $ User_{1} $ and $ User_{2} $ decode the desired information from the mixed information as mentioned in 2.2.

\section{Outage Analysis}

This section desires mathematical expressions of outage probability for PSR and TSR scheme one after another.

\subsection{Outage analysis of PSR }

The received signal at {\em EHR}, used to decode the information in $k^{th}$ slot, is given by

\begin{multline}
y_{k, r}^{PS} = \underbrace{\dfrac{\sqrt{\alpha_k{P_{k}^{T}}}h_{k}s_{k}^{PS}}{\sqrt{d_{k}^{\nu}}}}_\text{Signal}
+\underbrace{\dfrac{\sqrt{\alpha_kp}\beta_c s_{j}^{PS}}{\sqrt{d_{j}^{\nu}}}}_\text{Interference}\\+\underbrace{n_{r,at_k}+
n_{r,sc}}_\text{Noise},  (k \in 1,2)
\end{multline}
where, $ s_{k}^{PS}$ represents the signal of $User_{k}$ ($k \in 1,2$) with zero mean and variance E[$|s_{k}^{PS}|^{2}$]=1; $ s_{j}^{PS}$ indicates the signal from the interferer with zero mean and E[$|s_{j}^{PS}|^{2}$]=1; {\em p} and $d_j$ represent CCI power and distance between interferer and {\em EHR}, respectively; $n_{r,at_k}$ and $n_{r,sc}$ are used to indicate received antenna noise and RF to baseband signal conversion noise, respectively. In both the cases, noise is assumed to be additive white Gaussian noise (AWGN) with average power of $ \sigma_{r,at_k}^{2} $ and $ \sigma_{r,sc}^{2} $, respectively. Accordingly, the data rate of the link from $User_{k} $ to {\em EHR} can be expressed as 

\begin{equation}
R_{k, r}^{PS}=\scalebox{0.8}{$\dfrac{(T/4)}{T}$}\log_2{\scalebox{0.8}{$\Bigg(1+\dfrac{\alpha_{k}P_{k}^{T}|h_k|^{2}}{d_{k}^{\nu}\tiny {\bigg\{\sigma_{k,r}^{2}+{\dfrac{\alpha_{k}p|\beta_c|^{2}}{d_{j}^{\nu}}}}\bigg\}}\Bigg)$}},  (k \in 1,2)
\end{equation}
where, $\sigma_{k, r}^{2}=\sigma_{r,at_k}^{2}+\sigma_{r,sc}^{2}$. It is considered that RF energy circuit of the given system is incapable of harvesting energy from noise signal. Here, noise power of the signal available to the relay for decoding the information from {\em $User_k$} is assumed $\sigma_{k, r}^{2}$. 

Now the energy harvested at {\em EHR} from the transmissions in corresponding slots (in Figure 2) can be expressed as 

\begin{multline}
E_{k, r}^{PS}=\eta (1- \alpha_{k})\scalebox{0.8}{$\bigg(P_{k}^{R}+\dfrac{p|\beta_c|^{2}}{d_{j}^{\nu}}\bigg)\dfrac{T}{4}$}\\=\eta (1- \alpha_{k})\scalebox{0.8}{$\bigg(\dfrac{P_{k}^{T}|h_k|^{2}}{d_{k}^{\nu}}+\dfrac{p|\beta_c|^{2}}{d_{j}^{\nu}}\bigg)\dfrac{T}{4}$}, (k \in 1,2)
\end{multline}
where, $0 < \eta \leq 1$ defines the energy conversion efficiency, the value of which depends on the harvesting circuitry.

Based on the first two slots of transmission, harvested energy at {\em EHR} can be obtained as $\sum_{k=1}^{2} E_{k,r}^{PS} $. In time slot-3, this harvested energy is used to broadcast the coded information from {\em EHR} to the users. The received signal at destination node $User_k$ can be written as 

\begin{equation}
y_{rs_k}^{PS}=\dfrac{\sqrt{P_{r}^{PS}}{g_{k}} \hskip 0.1cm s^{PS}}{\sqrt{d_k^{\nu}}}+n_{s,k}+n_{s,c_k}, (k \in 1,2)
\end{equation}
where, {\em $s^{PS}$} is the coded version of $s_{1}^{PS}$ and $s_{2}^{PS}$; $n_{s,k}$ indicates AWGN between the path {\em EHR} to $User_k$ with an average power of $\sigma_{s,k}^2$ and $n_{s,c_k}$ indicates the RF to baseband signal conversion noise with variance of $\sigma_{s,c_k}^2$. Assuming {\em EHR} broadcasts using power $P_{r}^{PS}$, during T/2 interval, $P_{r}^{PS}$ can be expressed as follows 

$P_{r}^{PS}$=$\dfrac{\sum_{k=1}^{2} E_{k, r}^{PS}}{\dfrac{T}{2}}\\={\eta\sum_{k=1}^{2} \dfrac{P_{k}^{T}(1-\alpha_k) |h_{k}|^{2}}{2 d_{k}^{\nu}}} + \dfrac{\eta p |\beta_{c}|^2(2-\alpha_1-\alpha_2)}{2 d_{j}^{\nu}}$

\begin{equation}
= a |h_{1}|^{2} + b |h_{2}|^{2} + c |\beta_{c}|^2 = X (say),
\end{equation}
where, $a$ = \scalebox{0.8}{${\dfrac{\eta P_{1}^{T} (1-\alpha_1)} {2 d_{1}^{\nu}}}$}, $b$ = \scalebox{0.8}{$ {\dfrac{\eta P_{2}^{T} (1-\alpha_2)} {2 d_{2}^{\nu}}}$} and $c$ = \scalebox{0.8}{$\dfrac{\eta p(2-\alpha_1-\alpha_2)}{2 d_{j}^{\nu}}$}. 

As seen from (6), this harvesting power is a random variable, which is a linear weighted sum of three independent exponential functions. Hence, its pdf is expressed as follows [21] :

\begin{multline}
f_{X}(x)=\dfrac{\Omega_{h_{1}} \Omega_{h_{2}} \Omega_{\beta_{c}}}{a b c}\Bigg[\dfrac{e^{-\Big(\scalebox{0.75}{$\Omega_{h_{1}}\dfrac{x}{a}$}\Big)}}{\scalebox{0.75}{$\bigg(\dfrac{\Omega_{h_{2}}}{b}-\dfrac{\Omega_{h_{1}}}{a}\bigg)\bigg(\dfrac{\Omega_{\beta_{c}}}{c}-\dfrac{\Omega_{h_{1}}}{a}\bigg)$}}+\\\dfrac{e^{-\Big(\scalebox{0.75}{$\Omega_{h_{2}}{\dfrac{x}{b}}$}\Big)}}{\scalebox{0.75}{$\bigg(\dfrac{\Omega_{h_{1}}}{a}-\dfrac{\Omega_{h_{2}}}{b}\bigg)\bigg(\dfrac{\Omega_{\beta_{c}}}{c}-\dfrac{\Omega_{h_{2}}}{b}\bigg)$}}+\dfrac{e^{-\Big(\scalebox{0.75}{$\Omega_{\beta_{c}}\dfrac{x}{c}$}\Big)}}{\scalebox{0.75}{$\bigg(\dfrac{\Omega_{h_{2}}}{b}-\dfrac{\Omega_{\beta_{c}}}{c}\bigg)\bigg(\dfrac{\Omega_{h_{1}}}{a}-\dfrac{\Omega_{\beta_{c}}}{c}\bigg)$}}\Bigg]\\=\dfrac{1}{M}\Bigg[\dfrac{e^{\scalebox{0.75}{$-\Big(\Omega_{h_{1}}\dfrac{x}{a}\Big)$}}}{m}+\dfrac{e^{\scalebox{0.75}{$-\Big(\Omega_{h_{2}}\dfrac{x}{b}\Big)$}}}{n}+\dfrac{e^{\scalebox{0.75}{$-\Big(\Omega_{\beta_{c}}\dfrac{x}{c}\Big)$}}}{q}\Bigg]
\end{multline}
where, \scalebox{0.75}{$m = {{\bigg(\dfrac{\Omega_{h_{2}}}{b}-\dfrac{\Omega_{h_{1}}}{a}\bigg)\bigg(\dfrac{\Omega_{\beta_{c}}}{c}-\dfrac{\Omega_{h_{1}}}{a}\bigg)}}$}, \scalebox{0.75}{$ n = {\bigg(\dfrac{\Omega_{h_{1}}}{a}-\dfrac{\Omega_{h_{2}}}{b}\bigg)\bigg(\dfrac{\Omega_{\beta_{c}}}{c}-\dfrac{\Omega_{h_{2}}}{b}\bigg)}$} and  \scalebox{0.75}{$ q = {\bigg(\dfrac{\Omega_{h_{2}}}{b}-\dfrac{\Omega_{\beta_{c}}}{c}\bigg)\bigg(\dfrac{\Omega_{h_{1}}}{a}-\dfrac{\Omega_{\beta_{c}}}{c}\bigg)}$}; \scalebox{0.75}{$ M=\dfrac{a b c}{\Omega_{h_{1}} \Omega_{h_{2}} \Omega_{\beta_{c}}}$}.

The received data rate at $User_k$ from {\em EHR} can be given as 
\begin{equation}
R_{rs_k}^{PS}=\dfrac{1}{2}\log_2\Bigg(\scalebox{0.95}{$1+\dfrac{P_{r}^{PS}|g_{k}|^2}{d_{k}^\nu \sigma_{sc,k}^2}$}\Bigg), (k \in 1,2)
\end{equation} 
where, ${\sigma_{sc,k}^2}={\sigma_{s,k}^2}+{\sigma_{s,c_k}^2}, (k \in 1,2)$.

Due to the usage of digital NC, broadcast data rate is bounded by the performance of the weaken channel between $g_1$ and $g_2$, so real data rate of transmission in third slot is 
\begin{multline}
R_{bcrs}^{PS} = min\Big(R_{rs_k}^{PS}\Big) = \dfrac{1}{2}\log_2\Big(1+SNR_{bcrs}^{PS}\Big)
\\= \dfrac{1}{2}\log_2(1+P_{r}^{PS}L)
\end{multline}
where, $L=min\bigg(\scalebox{0.95}{$\dfrac{|g_{1}|^2}{d_{1}^\nu \sigma_{sc,1}^2}, \dfrac{|g_{2}|^2}{d_{2}^\nu \sigma_{sc,2}^2}$}\bigg)$. The complementary cumulative distribution function (CCDF) of {\em L} can be written as $F_L(l)=e^{\scalebox{0.8}{$-(d_1^{\nu}\sigma_{sc,1}^{2}\Omega_{g_{1}}+d_{2}^{\nu}\sigma_{sc,2}^{2}\Omega_{g_{2}})l$}}$ [22, Section 5.6].

As seen from Figure 1. this SWIPT-enabled NC based two-way {\em EHR} transmission network has three different transmission links, i.e., transmission between the links $User_1$ to {\em EHR}, $User_2$ to {\em EHR} and broadcast link of {\em EHR}. An outage is defined as the failure of any one of these links. Thus it is equivalent to complement of the event when transmission over all of these links meet their individual target rates. Hence, outage can be given as [23]

{$\scalebox{0.83}{$P_{out}^{PS}=1-(Pr[R_{1,r}^{PS}\geq R_1 ] \times Pr[R_{2, r}^{PS}\geq R_1 ]
Pr[R_{bcrs}^{PS}\geq R_1])$}$}

\begin{equation}  
    = 1-(1-P_{out,{U_{1}}R}^{PS})(1-P_{out,{U_{2}}R}^{PS} )(1-P_{out,BC}^{PS} )     
        \end{equation}
where, $R_1$ is the target rate of transmission in all the three different links.     
    	
    	Now, using (3), the outage probability between $User_k$ to {\em EHR} is defined as follows, 
    	 
\begin{equation}
P_{out,{U_{k}}R}^{PS}=1-Pr\bigg[\scalebox{0.75}{$\dfrac{\gamma_{h_k}^{PS}}{1+I_{R_k}^{PS}}$} \geq u_{k}^{PS}\bigg], (k \in 1,2)
\end{equation}
where, $ u_{k}^{PS} $ = $ 2^{(4R_1)}-1 $. The symbol $\gamma_{h_k}^{PS}=\scalebox{0.8}{$\dfrac{\alpha_kP_{k}^{T}}{d_k^{\nu}\sigma_{k, r}^2}|h_{k}|^{2}$}$ indicates the received SNR at relay from the transmission of $User_k$, while $I_{R_k}^{PS}=\scalebox{0.8}{$\dfrac{\alpha_kp}{\sigma_{k,r}^2 d_{j}^{\nu}} |\beta_{c}|^{2}$}$ indicates interference to noise ratio due to the CCI. Clearly, the distribution of $\gamma_{h_k}^{PS}$ is exponential in nature and it's probability density function (PDF) is given as 

\begin{equation}
f_{\gamma_{h_k}^{PS}}(v)=\scalebox{0.8}{$\dfrac{1}{\overline{\gamma}_{h_k}^{PS}}$}exp\bigg(-\scalebox{0.8}{$\dfrac{v}{\overline{\gamma}_{h_k}^{PS}}$}\bigg),  \thinmuskip=20mu  (v \geq 0)
\end{equation}
The distribution of $I_{R_k}^{PS}$ is also exponential in nature and it's probability density function (PDF) is given as 

\begin{equation}
f_{I_{R_k}^{PS}}(w)=\scalebox{0.8}{$\dfrac{1}{\overline\mu_{k}^{PS}}$}exp\bigg(-\scalebox{0.8}{$\dfrac{w}{\overline\mu_{k}^{PS}}$}\bigg), \thinmuskip=20mu   (w \geq 0)
\end{equation}
where, $ \overline{\gamma}_{h_k}^{PS}=\scalebox{0.8}{$\dfrac{\alpha_kP_{k}^{T}}{d_k^{\nu}\sigma_{k, r}^2}$}\Omega_{h_k}$ and $\overline\mu_{k}^{PS}=\scalebox{0.8}{$\dfrac{\alpha_kp}{\sigma_{k, r}^2 d_{j}^{\nu}}$}\Omega_{\beta_{c}}$. The symbols $\Omega_{h_k}$ and $ \Omega_{\beta_{c}} $ are the parameters of random variables $h_k$ and $ \beta_c $, respectively.

Now (11) can be rewritten as 

\begin{equation}
P_{out,{U_{k}}R}^{PS}=\dfrac{exp\bigg(\scalebox{0.8}{$\dfrac{1}{\overline\mu_{k}^{PS}}$}\bigg)}{\overline{\gamma}_{h_k}^{PS}\bigg(\scalebox{0.8}{$\dfrac{1}{\overline{\gamma}_{h_k}^{PS}}+\dfrac{1}{u_{k}^{PS}\overline\mu_{k}^{PS}}$}\bigg)}
\end{equation}

Appendix A provides the detail derivation of $P_{out,{U_{k}}R}^{PS}$.

On the other hand, using (9) outage probability between {\em EHR} to $User_k$ is 
\begin{multline}
P_{out,BC}^{PS}=1-Pr\Big[R_{bcrs}^{PS}\geq R_1\Big]\\=1-Pr\Big[\scalebox{0.8}{$\dfrac{1}{2}$}\log_{2}(1+SNR_{bcrs}^{PS})\geq R_1\Big]\\
=1-Pr[SNR_{bcrs}^{PS}\geq 2^{(2R_1)}-1]\\=1-Pr[SNR_{bcrs}^{PS}\geq u_{b}^{'}]\\=1-Pr[P_{r}^{PS} L\geq u_{b}^{'}]=1-Pr[XL\geq u_{b}^{'}]
\end{multline}
where, $u_{b}^{'}=2^{(2R_1)}-1$, X and L are given in (6) and (9), respectively.

Now using (6) and (7), $Pr[R_{bcrs}^{PS}\geq R_1]$ can be expressed as follows, 

\begin{multline}
Pr[R_{bcrs}^{PS}\geq R_1]=
\dfrac{2}{M}\Bigg\{\scalebox{0.8}{$\dfrac{\sqrt{{b_{o}u_{b}^{'}a}}}{m\sqrt{\Omega_{h_1}}}$}K_{1}\Bigg(\scalebox{0.8}{$2\sqrt{\dfrac{b_{o}u_{b}^{'}\Omega_{h_1}}{a}}$}\Bigg)\\
+\scalebox{0.8}{$\dfrac{\sqrt{{b_{o}u_{b}^{'}b}}}{n \sqrt{\Omega_{h_2}}}$}K_{1}\Bigg(\scalebox{0.8}{$2\sqrt{\dfrac{b_{o}u_{b}^{'}\Omega_{h_2}}{b}}$}\Bigg)
+\scalebox{0.8}{$\dfrac{\sqrt{{b_{o}u_{b}^{'}c}}}{q\sqrt{\Omega_{\beta_c}}}$}K_{1}\Bigg(\scalebox{0.8}{$2\sqrt{\dfrac{b_{o}u_{b}^{'}\Omega_{\beta_c}}{c}}$}\Bigg)\Bigg\}
\end{multline}

where, $b_o = (d_1^{\nu}\sigma_{sc,1}^{2}\Omega_{g_{1}}+d_{2}^{\nu}\sigma_{sc,2}^{2}\Omega_{g_{2}}) $

Detail derivation of this given mathematical expression of $Pr[R_{bcrs}^{PS}\geq R_1]$ is shown in Appendix B.

Finally, using (10), the equation of system outage probability in PSR (17) can be expressed as shown at the top of the next page (Page 7). where, $K_1$ indicates the first order modified Bessel function of second kind and it is defined as [24], {{$\int_{0}^{\infty}exp\bigg(-\scalebox{0.9}{$\dfrac{\lambda}{4t}-\gamma t\bigg)$}dt = \scalebox{0.9}{$\sqrt{\dfrac{\lambda}{\gamma}}$}K_1(\sqrt{\lambda\gamma})$}}

\begin{figure*}
\begin{equation}
\begin{split}
P_{out}^{PS}=1-\Bigg[\Bigg\{\scalebox{0.9}{$1-\dfrac{exp\bigg(\dfrac{1}{\overline\mu_{1}^{PS}}\bigg)}{\overline{\gamma}_{h_1}^{PS}\bigg(\dfrac{1}{\overline{\gamma}_{h_1}^{PS}}+\dfrac{1}{u_{1}^{PS}\overline\mu_{1}^{PS}}\bigg)}$}\Bigg\}\times
\Bigg\{\scalebox{0.9}{$1-\dfrac{exp\bigg(\scalebox{0.8}{$\dfrac{1}{\overline\mu_{2}^{PS}}$}\bigg)}{\overline{\gamma}_{h_2}^{PS}\bigg(\dfrac{1}{\overline{\gamma}_{h_2}^{PS}}+\dfrac{1}{u_{2}^{PS}\overline\mu_{2}^{PS}}\bigg)}$}\Bigg\}\times\dfrac{2}{M}\Bigg\{\scalebox{0.9}{$\dfrac{\sqrt{{b_{o}u_{b}^{'}a}}}{m\sqrt{\Omega_{h_1}}}$}K_{1}\Bigg(\scalebox{0.9}{$2\sqrt{\dfrac{b_{o}u_{b}^{'}\Omega_{h_1}}{a}}$}\Bigg)\\
+\scalebox{0.9}{$\dfrac{\sqrt{{b_{o}u_{b}^{'}b}}}{n \sqrt{\Omega_{h_2}}}$}K_{1}\Bigg(\scalebox{0.9}{$2\sqrt{\dfrac{b_{o}u_{b}^{'}\Omega_{h_2}}{b}}$}\Bigg)
+\scalebox{0.9}{$\dfrac{\sqrt{{b_{o}u_{b}^{'}c}}}{q\sqrt{\Omega_{\beta_c}}}$}K_{1}\Bigg(\scalebox{0.8}{$2\sqrt{\dfrac{b_{o}u_{b}^{'}\Omega_{\beta_c}}{c}}$}\Bigg)\Bigg\}
\Bigg]
\end{split}
\end{equation}
\hrulefill
\end{figure*}

\subsection{Outage analysis of TS Relaying scheme}
In TSR, the received signal at {\em EHR} from $ User_{k} $ to decode the information of $(k+1)^{th}$ slot is given by

\begin{multline}
y_{k,r}^{TS} = \underbrace{\dfrac{\sqrt{{P_{k}^{T}}}h_{k}s_{k}^{TS}}{\sqrt{d_{k}^{\nu}}}}_\text{Signal}+\underbrace{\dfrac{\sqrt{p}\beta_c}{\sqrt{d_{j}^{\nu}}} s_{j}^{TS}}_\text{Interference}\\+\underbrace{n_{r,at_k}+
n_{r,sc}}_\text{Noise},  (k \in 1,2)
\end{multline}
where, $ s_{k}^{TS}$ represents the signal of $User_{k}$ ($k \in 1,2$) with zero mean and variance E[$|s_{k}^{TS}|^{2}$]=1; $ s_{j}^{TS}$ indicates the signal from the interferer with zero mean and E[$|s_{j}^{TS}|^{2}$]=1. Similar to (3), the received data rate from $ User_{k} $ to {\em EHR} can be expressed as 

\begin{multline}
R_{k, r}^{TS}=\scalebox{0.8}{$\dfrac{\rho_l \dfrac{T}{2}}{T}$} \log_2\scalebox{0.8}{$\Bigg(1+\dfrac{P_{k}^{T}|h_k|^{2}}{d_{k}^{\nu}\bigg\{\sigma_{k, r}^{2}+\dfrac{p|\beta_c|^{2}}{d_{j}^{\nu}}\bigg\}}\Bigg)$},\\ (k \in 1,2; l \in 1,2)
\end{multline}

Applying TSR scheme, energy harvested at {\em EHR} from the transmissions of $User_{1}$ and $User_{2}$ can be expressed as 

\begin{multline}
E_{k, r}^{TS}=\eta \bigg(P_{k}^{R}+\scalebox{0.8}{$\dfrac{p|\beta_c|^{2}}{d_{j}^{\nu}}$}\bigg)\bigg(1-\sum_{l=1}^{2}\rho_l\bigg)\scalebox{0.8}{$\dfrac{T}{2}$}\\
       =\eta \bigg(\scalebox{0.8}{$\dfrac{P_{k}^{T}|h_k|^{2}}{d_{k}^{\nu}}$}+\scalebox{0.8}{$\dfrac{p|\beta_c|^{2}}{d_{j}^{\nu}}$}\bigg)\bigg(1-\sum_{l=1}^{2}\rho_l\bigg)\scalebox{0.8}{$\dfrac{T}{2}$}, (k \in 1,2)
\end{multline}

The total energy harvested at the {\em EHR} can be obtained as $\sum_{k=1}^{2} E_{k,r}^{TS} $. This harvesting energy is used to broadcast the coded information of {\em EHR} to $ User_k $. Similar to (5), the received signal at destination node $User_k$ can be written as 

\begin{equation}
y_{rs_k}^{TS}=\dfrac{\sqrt{P_{r}^{TS}}{g_{k}} \hskip 0.1cm s^{TS}}{\sqrt{(d_k)^{\nu}}}+n_{s,k}+n_{s,c_k}, (k \in 1,2)
\end{equation}

where, {\em $s^{TS}$} is the coded version of $s_{1}^{TS}$ and $s_{2}^{TS}$. Transmission power ($P_{r}^{TS}$) of {\em EHR} in fourth slot can be expressed as 

\begin{multline}
P_{r}^{TS}=\scalebox{0.8}{$\dfrac{\sum_{k=1}^{2} E_{k,r}^{TS}}{\dfrac{T}{2}}$}\\\scalebox{0.9}{$=\eta\dfrac{(1-\sum_{l=1}^{2}\rho_l)\sum_{k=1}^{2} { P_{k}^{T}  |h_{k}|^{2}}}{d_{k}^{\nu}}+ \dfrac{(1-\sum_{l=1}^{2}\rho_l)\eta p |\beta_{c}|^2}{d_{j}^{\nu}}$}\\=a^{'} h_{1}^{2}+b^{'} h_{2}^{2}+c^{'} \beta_{c}^{2}=Z (say),
\end{multline}
 where, $a^{'}=\scalebox{0.8} {$\dfrac{(1-\sum_{l=1}^{2}\rho_l)\eta P_{1}^{T}}{d_{1}^{\nu}}$}$, $b^{'}= \scalebox{0.8}{$\dfrac{(1-\sum_{l=1}^{2}\rho_l)\eta P_{2}^{T}}{d_{2}^{\nu}}$}$ and $c^{'}=\scalebox{0.8}{$\dfrac{(1-\sum_{l=1}^{2}\rho_l)\eta p}{d_{j}^{\nu}}$} $.

Received data rate at $User_k$ from {\em EHR} can be given as 

\begin{equation}
\begin{split}
R_{rs_k}^{TS}=\scalebox{1.0}{$\dfrac{T}{2T}$}\log_2\bigg(1+\scalebox{1.0}{$\dfrac{P_{r}^{TS}|g_{k}|^2}{d_{k}^\nu \sigma_{sc,k}^2}$}\bigg)\\=\scalebox{1.0}{$\dfrac{1}{2}$}\log_2\bigg(1+\scalebox{1.0}{$\dfrac{P_{r}^{TS}|g_{k}|^2}{d_{k}^\nu \sigma_{sc,k}^2}$}\bigg),
\end{split}
\end{equation}

Similar to the argument mentioned in (9), the data rate in third slot is expressed as 

\begin{equation}
R_{bcrs}^{TS} = min\big(R_{rs_k}^{TS}\big) = \scalebox{0.8}{$\dfrac{1}{2}$}\log_2\big(1+P_{r}^{TS}L^{'}\big),
\end{equation}
where, $L^{'}$ = min\bigg($\scalebox{0.95}{$\dfrac{|g_{1}|^2}{d_{1}^\nu \sigma_{sc,1}^2}, \dfrac{|g_{2}|^2}{d_{2}^\nu \sigma_{sc,2}^2}$}$\bigg). 

Similar to (10), the total outage probability can be written as  

{\scalebox{0.82}{$P_{out}^{TS}=1-(Pr[R_{1,r}^{TS}\geq R_1 ] \times Pr[R_{2,r}^{TS}\geq R_1 ] \times Pr[R_{bcrs}^{TS}\geq R_1])$}


\begin{equation}  
    = 1-(1-P_{out,{U_{1}}R}^{TS})(1-P_{out,{U_{2}}R}^{TS} )(1-P_{out,BC}^{TS} )     
\end{equation}
        
       Now, using (19), Outage probability between $User_k$ to {\em EHR} is   

 \begin{equation}
P_{out,{U_{k}}R}^{TS}=1-Pr\bigg[\scalebox{0.8}{$\dfrac{\gamma_{h_k}^{TS}}{1+I_{R_k}^{TS}}$} \geq u_{k}^{TS}\bigg]
\end{equation}
where, $ u_{k}^{TS} $ =\bigg \{$2^{\bigg(\scalebox{0.8}{$\dfrac{2 R_1}{\rho_l}$}\bigg)}-1\bigg\} $, $R_1$ is the target rate of transmission; $\gamma_{h_k}^{TS}=\scalebox{0.8}{$\dfrac{P_{k}^{T}}{\sigma_{k, r}^2(d_k)^{\nu}}$}|h_{k}|^{2}$ indicates the received SNR at relay due to the transmission of $User_k$. Here, $I_{R_k}^{TS}=\scalebox{0.8}{$\dfrac{p}{\sigma_{k,r}^2 (d_{j})^{\nu}}$}|\beta_{c}|^{2}$ indicates the interference to noise ratio (INR) due to the CCI. Both SNR and INR are random variables and follow exponential distribution with parameter $\overline{\gamma}_{h_{k}}^{TS}$ and $\overline\mu_{k}^{TS}$. The distribution of $\gamma_{h_k}^{TS}$ and $I_{R_k}^{TS}$ are similar to the distribution of (12) and (13), respectively; where, $ \overline{\gamma}_{h_{k}}^{TS}=\scalebox{0.8}{$\dfrac{P_{k}^{T}}{d_k^{\nu}\sigma_{k, r}^2}$}\Omega_{h_k}$, $\overline\mu_{k}^{TS}=\scalebox{0.8}{$\dfrac{p}{\sigma_{k, r}^2 d_{j}^{\nu}}$}\Omega_{\beta_{c}}$. 

Similar to (14), now (26) can be rewritten as 
\begin{equation}
P_{out,{U_{k}}R}^{TS}=\dfrac{exp\bigg(\scalebox{0.8}{$\dfrac{1}{\overline\mu_{k}^{TS}}$}\bigg)}{\overline{\gamma}_{h_k}^{TS}\bigg(\scalebox{0.8}{$\dfrac{1}{\overline{\gamma}_{h_k}^{TS}}+\dfrac{1}{u_{k}^{TS}\overline\mu_{k}^{TS}}$}\bigg)}, (k \in 1,2)
\end{equation}

Outage probability between {\em EHR} to $User_k$ using (25) is expressed as

\begin{equation}
P_{out,BC}^{TS}=1-Pr\big[R_{bcrs}^{TS}\geq R_1\big]
=1-Pr\big[L^{'}Z\geq u_{t}^{'}\big]
\end{equation}

where, $u_{t}^{'}= 2^{\scalebox{0.8}{$(2R_1)$}}-1$.

Similar to (16), $Pr[R_{bcrs}^{TS}\geq R_1]$ can be written as 

\begin{multline}
Pr\big[R_{bcrs}^{TS}\geq R_1\big]= Pr\big[L^{'}Z\geq u_{t}^{'}\big]\\
=\dfrac{2}{M^{'}}\Bigg\{\scalebox{0.8}{$\dfrac{\sqrt{{b_{o}u_{t}^{'}a^{'}}}}{m^{'}\sqrt{\Omega_{h_1}}}$}K_{1}\Bigg(\scalebox{0.8}{$2\sqrt{\dfrac{b_{o}u_{t}^{'}\Omega_{h_1}}{a^{'}}}$}\Bigg)+\\
\scalebox{0.8}{$\dfrac{\sqrt{{b_{o}u_{t}^{'}b^{'}}}}{n^{'}\sqrt{\Omega_{h_2}}}$}K_{1}\Bigg(\scalebox{0.8}{$2\sqrt{\dfrac{b_{o}u_{t}^{'}\Omega_{h_2}}{b^{'}}}$}\Bigg)+
\scalebox{0.8}{$\dfrac{\sqrt{{b_{o}u_{t}^{'}c^{'}}}}{q^{'}\sqrt{\Omega_{\beta_c}}}$}K_{1}\Bigg(\scalebox{0.8}{$2\sqrt{\dfrac{b_{o}u_{t}^{'}\Omega_{\beta_c}}{c^{'}}}$}\Bigg)\Bigg\}
\end{multline}

\begin{figure*}
\hrulefill
\begin{equation}
\begin{split}
P_{out}^{TS}=1-\Bigg[\Bigg\{\scalebox{0.9}{$1-\dfrac{exp\bigg(\dfrac{1}{\overline\mu_{1}^{TS}}\bigg)}{\overline{\gamma}_{h_1}^{TS}\bigg(\dfrac{1}{\overline{\gamma}_{h_1}^{TS}}+\dfrac{1}{u_{1}^{TS}\overline\mu_{1}^{TS}}\bigg)}$}\Bigg\}\times
\Bigg\{\scalebox{0.9}{$1-\dfrac{exp\bigg(\dfrac{1}{\overline\mu_{2}^{TS}}\bigg)}{\overline{\gamma}_{h_2}^{TS}\bigg(\dfrac{1}{\overline{\gamma}_{h_2}^{TS}}+\dfrac{1}{u_{2}^{TS}\overline\mu_{2}^{TS}}\bigg)}$}\Bigg\}\times
\dfrac{2}{M^{'}}\Bigg\{\scalebox{0.9}{$\dfrac{\sqrt{{b_{o}u_{t}^{'}a^{'}}}}{m^{'}\sqrt{\Omega_{h_1}}}$}K_{1}\Bigg(\scalebox{0.9}{$2\sqrt{\dfrac{b_{o}u_{t}^{'}\Omega_{h_1}}{a^{'}}}$}\Bigg)+\\
\scalebox{0.9}{$\dfrac{\sqrt{{b_{o}u_{t}^{'}b^{'}}}}{n^{'}\sqrt{\Omega_{h_2}}}$}K_{1}\Bigg(\scalebox{0.9}{$2\sqrt{\dfrac{b_{o}u_{t}^{'}\Omega_{h_2}}{b^{'}}}$}\Bigg)
+\scalebox{0.9}{$\dfrac{\sqrt{{b_{o}u_{t}^{'}c^{'}}}}{q^{'}\sqrt{\Omega_{\beta_c}}}$}K_{1}\Bigg(\scalebox{0.9}{$2\sqrt{\dfrac{b_{o}u_{t}^{'}\Omega_{\beta_c}}{c^{'}}}$}\Bigg)\Bigg\}\Bigg]
\end{split}
\end{equation}
\hrulefill
\end{figure*}

where, \scalebox{0.75}{$M^{'}=\dfrac{a^{'} b^{'} c^{'}}{\Omega_{h_{1}} \Omega_{h_{2}} \Omega_{\beta_{c}}}$},
 \scalebox{0.75}{$m^{'} = {\bigg({\dfrac{\Omega_{h_{2}}}{b^{'}}-\dfrac{\Omega_{h_{1}}}{a^{'}}\bigg)\bigg(\dfrac{\Omega_{\beta_{c}}}{c^{'}}-\dfrac{\Omega_{h_{1}}}{a^{'}}\bigg)}}$}, 
\scalebox{0.7}{$n^{'} = {\bigg(\dfrac{\Omega_{h_{1}}}{a^{'}}-\dfrac{\Omega_{h_{2}}}{b^{'}}\bigg)\bigg(\dfrac{\Omega_{\beta_{c}}}{c^{'}}-\dfrac{\Omega_{h_{2}}}{b^{'}}\bigg)}$}, \scalebox{0.7}{$q^{'} = {\bigg(\dfrac{\Omega_{h_{2}}}{b^{'}}-\dfrac{\Omega_{\beta_{c}}}{c^{'}}\bigg)\bigg(\dfrac{\Omega_{h_{1}}}{a^{'}}-\dfrac{\Omega_{\beta_{c}}}{c^{'}}\bigg)}$}. 

Finally, outage expression of TSR (30) can be obtained as shown at the top of the next page (Page 8).

\subsection{Spectrum Efficiency and Energy Efficiency}

Results of the outage analysis is used to evaluate spectrum-efficiency (or throughput)
and energy-efficiency of the proposed system in presence of CCI and performance comparison is done over the similar system without CCI. These two metrics can be defined as follows :

\begin{equation}
\begin{split}
\eta_{SE}^{PS}=(1-P_{out}^{PS})*R_1*2*\dfrac{T}{2T},       (for PSR) \\    
\eta_{SE}^{TS}=(1-P_{out}^{TS})*R_1*2*\dfrac{T}{2T},       (for TSR)  
\end{split} 
\end{equation}

\begin{equation}
\begin{split}
\eta_{EE}^{PS}=\dfrac{\eta_{SE}^{PS}}{Transmit \, power},       (for PSR) \\    
\eta_{EE}^{TS}=\dfrac{\eta_{SE}^{TS}}{Transmit \, power},       (for TSR)  
\end{split} 
\end{equation}

Evaluation of the performance of proposed system using these metrics is discussed in the next section.

\section{Simulation Results}
 
In this section, first we present the performance results of simulation study to validate our analytical results derived in the previous section. Further, the results are also shown to illustrate the variation in system outage probability for the bidirectional communication with the key system parameters; (i) power splitting ratio, (ii) time switching factors and (iii) transmitted signal power. Here, normalized transmitted signal power is considered as the ratio of transmitted power to noise power. Due to low noise power, the value of this normalized transmitted signal power in dB is very high. Performance of PSR and TSR schemes are compared with that of a similar system without CCI [14]. Results of comparative system model without CCI is generated considering PSR scheme with three equal time slot durations. Values of necessary system parameters are listed in Table II. Additionally, the distance between two users is considered as 14 m. 
\begin{table}\caption{List of necessary parameters}
  \centering
    \begin{tabular}{l|p{15mm}}
    \hline
    Name & Value\\
    \hline
    \hline
     Target rate of transmission ($R_{1}$) & 0.9 bps/Hz \\
    \hline
     Path loss exponent ($ \nu $)   & 2.7 (for urban area) \\  
    \hline
     Noise components, \\ \newline ${\sigma_{1, r}^{2}}={\sigma_{2, r}^{2}}
     =     {\sigma_{sc, 1}^{2}}={\sigma_{sc, 2}^{2}}$    & 1 $\mu W$ \\
    \hline
     Time switching factor, $\rho_1 = \rho_2 = \rho$    &    0.32\\
    \hline
     Power splitting factor, $\alpha_1 = \alpha_2 = \alpha$    &    0.14\\
     \hline
     Energy conversion efficiency, $\eta $    &    0.8\\
    \hline
     \end{tabular}
  \end{table}
  
  \begin{figure}
  \centering{
\includegraphics[scale=0.56]{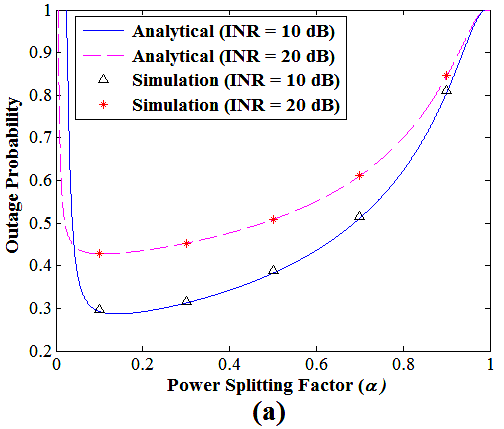}
\includegraphics[scale=0.56]{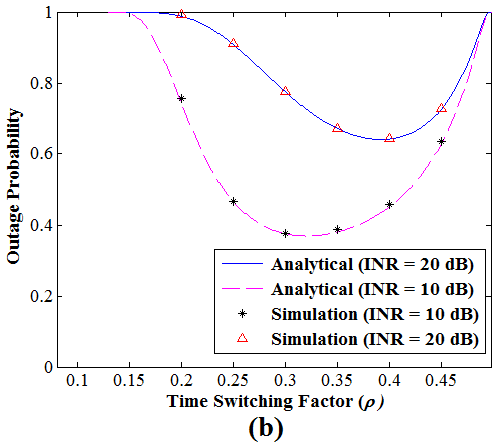}
\includegraphics[scale=0.564]{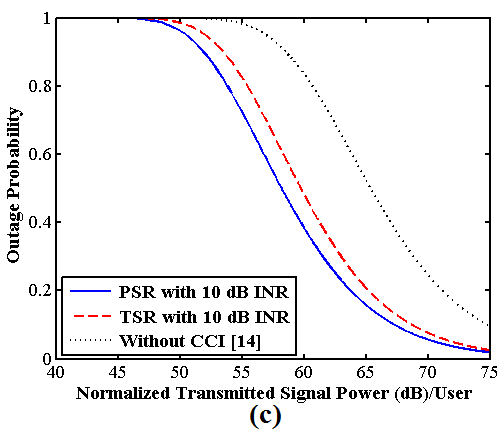} }
\caption {(a) Graphical plot of outage probability versus power splitting factor ($\alpha$) with transmitted power 1.5 Watt, (b) Graphical plot of outage probability versus time switching factor ($\rho$) with transmitted power 1.5 Watt, (c) Outage probability versus normalized transmitted signal power with $\alpha=0.14$ and $\rho=0.32$}  \label{fig:img_2}
\end{figure}

Figure 4.a. depicts the system outage performance versus power splitting factor ($\alpha$), assuming only PSR based SWIPT operation in the system. It is clearly seen that the theoretical result shown in (17), closely matches with the simulation results. The outage performance is very poor when $\alpha$ is very high, almost close to 1 and the outage probability decreases with decrease in the value of $\alpha$. This may be explained as follows: as $\alpha$ increases, more power is allocated for user information transmission and less power is used for energy harvesting, following (3) and (4). Due to this insufficient harvesting energy, relay broadcasting rate is more likely to fail to meet the target rate of information transmission. With the decrease in $\alpha$, the system outage probability decreases till the optimal value of $\alpha_{opt}$ is reached. When $\alpha=\alpha_{opt}=0.14$, the harvesting power at {\em EHR} is found to be sufficient. Thus the achievable rate of the broadcast link; more likely satisfies the target rate of information transmission. For $0<\alpha<\alpha_{opt}$, inadequate power allocation for information decoding in presence of noise (poor SNR), increases the system outage probability.

Figure 4.b. shows the system's outage probability versus time switching factor ($\rho$), assuming only TSR based SWIPT operation in the system. When $\rho$ is very small, the time allocation for information transmission is less and the time allocation for energy harvesting is more, following (19) and (20). Due to this insufficient slot duration, the average achievable rate is poor (19) and thus it is more likely to fail to meet the target rate of transmission. This results in high outage of links from $User_k$ to {\em EHR} ($k\in1,2$), as shown in figure. With the increase in $\rho$ value, the system outage probability decreases due to SNR improvement of the links from $User_k$ to {\em EHR}, following (19), (27), (30) and reaches the minimum value at the optimal value of $\rho_{opt} = 0.32$. However, when $\rho$ value exceeds the optimal value, inadequate harvesting energy fails to meet the target SNR of relay broadcast and effectively the system outage probability is gradually increased with the increasing value of $\rho$. 

As CCI increases (from INR 10 dB to 20 dB) the outage performance of the system becomes worse. This may be observed both in Figure 4.a. and 4.b. This is due to the fact that SNR of the links from $User_k$ to {\em EHR} ($k\in1,2$) become worse due to high interference power. As a result these two together make overall outage probability high.

Figure 4.c. displays the system outage performance vs normalized transmitted signal power for PSR and TSR schemes in presence of CCI to have a comparison with that of a similar system without CCI. Initially, the outage falls rapidly as the normalized transmitted signal power increases in the range of 54-65 dB (55-63 dB) for PSR (TSR) scheme. About 82$\%$ and 75$\%$ less power are required to achieve the outage probability 0.2 for PSR and TSR schemes, respectively in presence of CCI as compared to the system without CCI. Similarly, the amount of less power is observed as 82.5$\%$ and 76$\%$ for outage probability 0.01. The reason of this nature can be explained as follows: when transmitted power is less, low power allocation for information transmission as well as for energy harvesting could not satisfy the target rate of the links between $User_k$ ($k\in1,2$) and the {\em EHR}. With the increase in normalized transmitted signal power, SNR is improved and the outage probability drops. 

\begin{figure}
\includegraphics[scale=0.44]{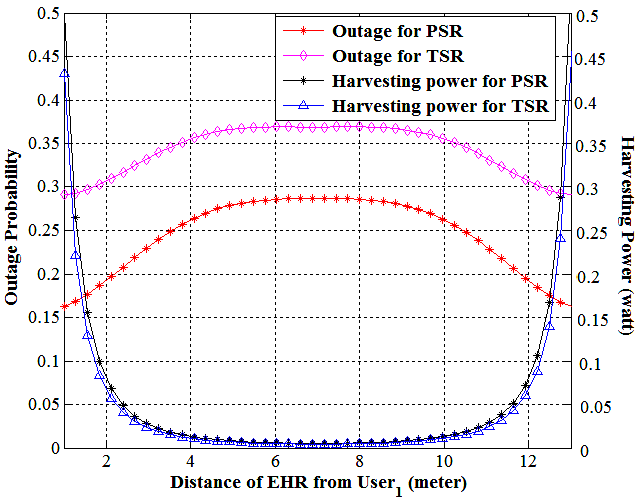}
\caption {Graphical plot of outage probability and harvesting power versus relay position with transmitted power 1.5 Watt, INR=10 dB, $\alpha=0.14$ and $\rho=0.32$}
\end{figure}

\begin{figure}
\includegraphics[scale=0.55]{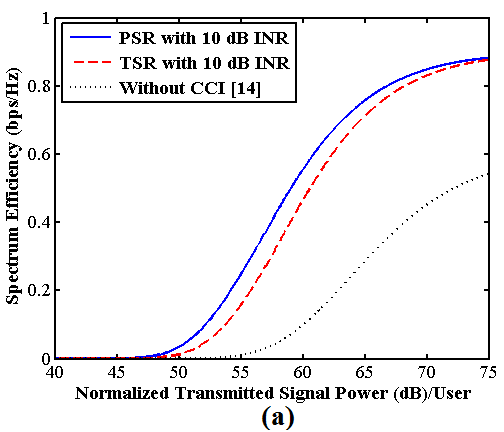}
\includegraphics[scale=0.53]{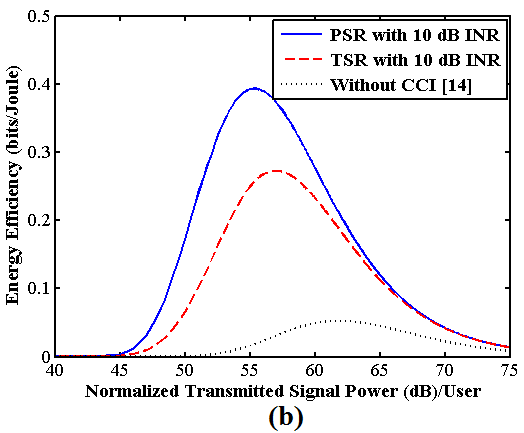}
\caption {(a) Spectrum efficiency versus normalized transmitted signal power with $\alpha=0.14$ and $\rho=0.32$,
(b) Energy efficiency versus normalized transmitted signal power with $\alpha=0.14$ and $\rho=0.32$}
\end{figure}

Figure 5. illustrates the variation in the outage probability of both PSR and TSR, respectively, with respect to the distance of the {\em EHR} from $user_1$. This figure shows the importance of relay placement on the outage performance for both PSR and TSR schemes. When relay moves away from $user_1$ or, $user_2$, outage increases initially but then it remains flat for a certain range of relay position. The flat outage probability is observed in the range of 5.5-8.5 m (5.8-8.2 m) distant from $user_1$, for TSR (PSR). To explain this nature, the variation of harvesting power in two-way relaying protocol is also shown in the same figure. When relay is placed closer to user, it can harvest more power for broadcast as compared to the relay placement at the middle of $user_1$ and $user_2$, as shown in figure. On the other hand, the product of two outage probabilities between the links $User_k$ ($k\in1,2$) and {\em EHR} are comparatively less when the {\em EHR} is placed closer to one of the users as compared to the relay position at the middle of $user_1$ and $user_2$, following (14) or, (27). About 22$\%$ improved outage is observed for the worst case performance at PSR as compared to TSR scheme. 

Figure 6.a. presents the results of spectrum efficiency in PSR and TSR schemes for CCI = 10 dB and performance comparison is done with that of a similar system without CCI. The nature of this plot of spectrum efficiency is just opposite of the graphical plot of outage probability as shown in Figure 4.c. The rapid enhancement of spectrum efficiency is observed in the range of 52-61 dB  and 54-60 dB normalized signal power for PSR and TSR schemes, respectively. As the normalized signal power increases beyond the critical value (75 dB is observed), the improvement of spectrum efficiency is negligible as compared to the two-way relaying scheme without CCI. About 86$\%$ and 80$\%$ less power are required to achieve the spectrum efficiency 0.2 bps/Hz for PSR and TSR schemes, respectively in presence of CCI as compared to the system without CCI. Similarly, the amount of less power is observed as 95$\%$ and 93$\%$ to achieve the spectrum efficiency 0.5 bps/Hz. 

Figure 6.b. depicts the energy efficiency in PSR and TSR schemes with CCI and performance comparison is done with that of a similar system without CCI. As seen from the figure, if the transmitted power increases the energy efficiency also increases accordingly. This is due to the fact that the transmitted signal power increases till an optimal value (55.3 dB for PSR and 57 dB for TSR), the non-linear change of the spectrum efficiency is more as compared to the linear change of signal power which increases the energy efficiency. Since the harvested energy proportionally increases with the transmitted power, this harvesting energy can improve the relaying ability in broadcast phase. Eventually, outage probability drops for increasing value of the transmitted signal power till the optimal value. Effectively, the spectrum efficiency increases due to better outage performance and the energy efficiency improves. The energy efficiency reaches its peak at an optimal value of signal power. Further if the signal power increases beyond the optimal value, the change in the spectrum efficiency value is less than the change in signal power which decreases the energy efficiency. PSR (TSR) scheme is about 87$\%$ (81$\%$) more energy efficient as compared to the system without CCI in two-way communication. 

\section{Conclusions}
  
In this paper, an analytical study of outage performance is carried out for two-way communication in a SWIPT enabled DF relay assisted network in presence of CCI. Closed form expressions of outage probability are derived for both PSR and TSR schemes in terms of system parameters like relay positioning, time switching factor, power splitting factor. Simulation results highlight the closeness to the analytical results. It is shown that in terms of outage probability, spectrum and energy efficiency, two-way RF-EH in presence of CCI is found to be more efficient than the system operates without CCI. The present system model may be extended as a outage minimization problem on multiple antenna cognitive radio network in fully energy causality system with an interference constraint.

\section*{Appendix}
\subsection*{Derivation of $P_{out,{U_{k}}R}^{PS}$}



This appendix provides the detailed derivation of $P_{out,{U_{k}}R}^{PS}$ for PSR scheme. The $P_{out,{U_{k}}R}^{TS}$ for TSR scheme follows the same procedure as provided below.
 
\begin{multline}
P_{out,{U_{k}}R}^{PS}=1-Pr\bigg[\scalebox{0.8}{$\dfrac{\gamma_{h_k}^{PS}}{1+I_{R_k}^{PS}}$} \geq u_{k}^{PS}\bigg], (k \in 1,2)\\
=1-\int_{0}^{\infty}Pr\Big[I_{R_k}^{PS}\leq \scalebox{0.8}{$\dfrac{\gamma_{h_k}^{PS}}{u_{k}^{PS}}$}-1\Big]f_{\gamma_{h_k}^{PS}}(v) dv\\
=1-\int_{0}^{\infty}\Bigg[\scalebox{0.8}{$\dfrac{1}{\overline\mu_{k}^{PS}}$}\int_{0}^{\scalebox{0.7}{$\Big(\dfrac{v}{u_{k}^{PS}}-1\Big)$}}exp\Big(-\scalebox{0.85}{$\dfrac{w}{\overline\mu_{k}^{PS}}$}\Big)dw\Bigg]f_{\gamma_{h_k}^{PS}}(v) dv\\
=1+\scalebox{0.8}{$\dfrac{1}{\overline{\gamma}_{h_k}^{PS}}$}\int_{0}^{\infty}\bigg[exp\bigg(-\scalebox{0.8}{$\dfrac{\dfrac{v}{u_{k}^{PS}}-1}{\overline\mu_{k}^{PS}}$}\bigg)-1\bigg]exp\bigg(-\scalebox{0.8}{$\dfrac{v}{\overline{\gamma}_{h_k}^{PS}}$}\bigg)dv\\
=\dfrac{exp\bigg(\scalebox{0.8}{$\dfrac{1}{\overline\mu_{k}^{PS}}$}\bigg)}{\overline{\gamma}_{h_k}^{PS}\bigg(\scalebox{0.8}{$\dfrac{1}{\overline{\gamma}_{h_k}^{PS}}+\dfrac{1}{u_{k}^{PS}\overline\mu_{k}^{PS}}$}\bigg)}
\end{multline}

\subsection*{Derivation of 16}
 
Same mathematical steps are followed in (29).
\begin{multline}
Pr[R_{bcrs}^{PS}\geq R_1]=Pr[XL\geq u_{b}^{'}]
\\=\int_{0}^{\infty}Pr\Big[L\geq \scalebox{0.8}{$\dfrac{u_{b}^{'}}{X}$}\Big].f_X(x)dx\\
=\int_{0}^{\infty}e^{-\scalebox{0.8}{$(d_1^{\nu}\sigma_{sc,1}^{2}\Omega_{g_{1}}+d_{2}^{\nu}\sigma_{sc,2}^{2}\Omega_{g_{2}})\dfrac{u_{b}^{'}}{x}$}}.f_X(x)dx\\
=\dfrac{2}{M}\Bigg\{\scalebox{0.8}{$\dfrac{\sqrt{{b_{o}u_{b}^{'}a}}}{m\sqrt{\Omega_{h_1}}}$}K_{1}\Bigg(\scalebox{0.8}{$2\sqrt{\dfrac{b_{o}u_{b}^{'}\Omega_{h_1}}{a}}$}\Bigg)
+\\\scalebox{0.8}{$\dfrac{\sqrt{{b_{o}u_{b}^{'}b}}}{n \sqrt{\Omega_{h_2}}}$}K_{1}\Bigg(\scalebox{0.8}{$2\sqrt{\dfrac{b_{o}u_{b}^{'}\Omega_{h_2}}{b}}$}\Bigg)
+\scalebox{0.8}{$\dfrac{\sqrt{{b_{o}u_{b}^{'}c}}}{q\sqrt{\Omega_{\beta_c}}}$}K_{1}\Bigg(\scalebox{0.8}{$2\sqrt{\dfrac{b_{o}u_{b}^{'}\Omega_{\beta_c}}{c}}$}\Bigg)\Bigg\}
\end{multline}

\end{document}